# Phase Behavior and Mesoscale Solubilization in Aqueous Solutions of Hydrotropes

Deepa Subramanian and Mikhail A. Anisimov*

Department of Chemical & Biomolecular Engineering, University of Maryland, College Park, MD 20742, USA
Institute for Physical Science & Technology, University of Maryland, College Park, MD 20742, USA

*Corresponding author: anisimov@umd.edu

**Abstract**

Hydrotropes are amphiphilic molecules that are too small to spontaneously form equilibrium structures in aqueous solutions, but form dynamic, non-covalent assemblies, referred to as clusters. In the presence of a hydrophobic compound, these clusters seem to get stabilized leading to the formation of long-lived, highly stable mesoscopic droplets, a phenomenon that we call "mesoscale solubilization". In this work, we focus on the unusual mesoscopic properties of aqueous solutions of a nonionic hydrotrope, namely tertiary butyl alcohol (TBA), on addition of various hydrophobic compounds. Aqueous TBA solutions, in about 3 mol % to 8 mol % TBA concentration range and about 0 °C to 25 °C temperature range, show the presence of short-ranged (~ 0.5 nm), short-lived (tens of picoseconds) molecular clusters which result in anomalies of the thermodynamic properties. These clusters are transient but do not relax by diffusion, thus distinctly different from conventional concentration fluctuations. In this concentration and temperature range, upon the addition of a third (more hydrophobic) component to TBA-water solutions, long-lived mesoscopic droplets of about 100 nm size are observed. In this work, we clarify the ambiguity behind the definition of solubility and elucidate the phenomenon of mesoscale solubilization. A systematic study of the macroscopic and mesoscopic phase behavior of three ternary systems TBA-water-propylene oxide, TBA-water-isobutyl alcohol, and TBA-water-cyclohexane has been carried out. We differentiate between molecular solubility, mesoscale solubilization, and macroscopic phase separation. We have confirmed that practically stable aqueous colloids can be created from small molecules, without addition of surfactants or polymers. Such kind of novel materials may find applications in the design of various processes and products, ranging from pharmaceuticals to cosmetics and agrochemicals.

1. **Introduction**

Hydrotropes are small organic molecules that are known to increase the solubility of hydrophobic compounds in aqueous solutions [1]. Hydrotropes are amphiphilic molecules whose non-polar part is smaller when compared with traditional surfactants [2]. In an aqueous environment, hydrotropes do not spontaneously self-assemble to form stable equilibrium structures, such as micelles [3]. However, above a minimum hydrotrope concentration, they show the presence of a loose, dynamic non-covalent clustering or "micelle-like" fluctuations [4-7]. These micelle-like fluctuations are short-ranged, less than 1 nm in size and short-lived with a lifetime of tens of picoseconds [5,7-9]. Hydrotropes are classified as ionic and non-ionic hydrotropes, with aromatic ionic hydrotropes, such as sodium benzene sulphonate or sodium benzoate, being traditionally used in the industry [10]. Although there is a wide amount of literature on ionic hydrotropes, the role of the micelle-like fluctuations in aqueous solutions of non-ionic hydrotropes is rarely discussed. Examples of non-ionic hydrotropes include aromatic alcohols such as resorcinol, and amides such as urea [11,12]. In this work, we extend the definition of non-ionic hydrotropes to include short-chain alcohols such as *n*-propanol, *iso*-propanol, tertiary butyl alcohol, 2-butoxyethanol, amines such as 3-methylpyridine, and ethers such as tetrahydrofuran.

This work focuses on the properties of a specific non-ionic hydrotrope, namely tertiary butyl alcohol (TBA). TBA is the highest molecular weight alcohol to be completely miscible with water in all proportions under ambient conditions [13]. Recently, molecular dynamics simulations have shown the presence of micelle-like fluctuations over a dilute concentration range, 3 mol % to 8 mol % (11 mass % to 26 mass %) TBA in water [7]. Numerous experimental studies and other simulations support this picture [14-20]. Micelle-like structural fluctuations lead to anomalies in the thermodynamic properties of TBA-water solutions [7]. Thermodynamic properties such as activity coefficients, chemical potentials, and heat capacity, show extrema in this concentration range and at temperatures below room temperature [21-27].

Furthermore, under certain conditions aqueous TBA solutions show the presence of mesoscopic droplets, with a size of about 100 nm [28-35]. We have recently shown that mesoscopic droplets form only when a hydrophobic compound ("hydrophobe") is added to aqueous TBA solutions in the concentration range of 3 mol % to 8 mol % TBA and at low temperatures (below room temperature), thus clarifying the long-standing issue on the origin of mesoscopic inhomogeneities in such solutions [34,35]. The hydrophobe seems to stabilize the micelle-like fluctuations leading to formation of the mesoscopic droplets. We call the phenomenon of formation of mesoscopic droplets in aqueous solutions of hydrotropes with addition of a hydrophobe, the mesoscale solubilization.

Mesoscale solubilization is not unique to aqueous TBA solutions, but has been observed across various hydrotrope-water solutions that contain hydrophobic compounds. Yang *et al.* have observed the presence of mesoscopic droplets, 100s of nm in size, in aqueous solutions of tetrahydrofuran and 1,4-dioxane [36]. Sedlák has carried out extensive static and dynamic light scattering experiments in around 100 different solute-solvent pairs and observed the presence of mesoscopic droplets in aqueous solutions of various electrolytic and non-electrolytic solutions [37-39]. Jin *et al.* carried out a static and dynamic light scattering study in aqueous solutions of tetrahydrofuran, ethanol, urea, and α-cyclodextrin and observed mesoscopic droplets in these solutions, which they interpreted as gaseous nanobubbles [40]. In our previous work, we have clarified that such observed mesoscopic droplets in aqueous solutions of non-ionic hydrotropes are caused by the contamination of the non-ionic hydrotrope by trace amounts of hydrophobic impurities [34,35].

In this paper, we elucidate the phenomenon of mesoscale solubilization. We aim to clarify the ambiguity of the conventional definition of solubility by differentiating between molecular solubility and molecular clustering on the one hand, and macrophase separation and mesoscale solubilization on the other hand. We define mesoscale solubilization as the formation of mesoscopic droplets (order of a hundred nm in size) that leads to increased effective solubility of hydrophobic compounds in aqueous solutions of non-ionic hydrotropes. The thermodynamic stability of mesoscopic droplets in aqueous solutions of TBA in the presence of several hydrophobic compounds is discussed. Ternary phase behavior in three different systems has been studied: 1) TBA-water-propylene oxide; 2) TBA-water-isobutyl alcohol; 3) TBA-water-cyclohexane. In addition to determining the macroscopic ternary phase diagrams, the macroscopically homogeneous one-phase regions have also been characterized, via static and dynamic light scattering, to clarify the nature of the mesoscale solubilization.

2. **Experimental Section**

   *2.1 Materials*

   TBA with a labeled purity of 0.998+ was purchased from Alfa Aesar. Propylene oxide (PO) with a labeled purity of 0.995+ was purchased from Sigma Aldrich. Isobutyl alcohol (IBA) with a labeled purity of 0.999+ was purchased from J. T. Baker. Cyclohexane (CHX) with a labeled purity of 0.990+ was purchased from Merck. Deionized water was obtained from a Millipore setup.
   For light scattering experiments, the binary TBA-water solutions were filtered with 200 nm Nylon filters to remove dust particles. An additional filtration with 20 nm Anopore filters was carried out only if the TBA-water solutions showed mesoscopic droplets due to the presence of hydrophobic impurities in TBA (as was shown in ref. 35). The third component, PO, IBA, or CHX was added to the filtered TBA-water solution. PO was used without filtration because of its high volatility (boiling point 34 °C [41]), while IBA and CHX were used after filtering them with 200 nm Nylon filters. Light scattering measurements were performed after equilibrating the samples for about 24 hours.

The refractive index was measured with an Abbe refractometer at room temperature. The viscosity of the samples was measured with a Ubbelohde viscometer in a temperature controlled (± 0.2 °C) water bath.

*2.2 Determination of Phase Diagram*

The ternary phase diagram for each of the three systems was determined by the cloud-point method [42]. The third component was added to a binary mixture in small steps. At each step, the ternary mixture was manually shaken and let to rest for about 3 to 5 minutes. The sample was then visually observed to determine if phase transition had occurred. If not, more of the third component was added and the above procedure was repeated. The ternary phase diagram of TBA-water-PO system was determined at 25 °C, with an accuracy of ± 0.1 °C. The phase diagrams of TBA-water-IBA and TBA-water-CHX were determined at 21 °C, with an accuracy of ± 0.5 °C. In order to estimate the location of the critical point, light scattering experiments were carried out in the macroscopic one-phase region close to the binodal curve. If the correlation length of critical fluctuations exhibited a maximum, then the point of the binodal curve corresponding to this maximum was interpreted as the critical point.

*2.3 Light Scattering*

Static and dynamic light scattering experiments were performed with a PhotoCor Instruments setup, as described in ref. [34]. Temperature was controlled with an accuracy of ± 0.1 °C. The Photocor correlator computes an intensity auto-correlation function $g_2(t)$, which is an instrumental output that can be further analyzed (as shown below) to determine the size of aggregates undergoing Brownian motion in a system at thermodynamic equilibrium. For two exponentially decaying relaxation processes, auto-correlation function $g_2(t)$, obtained in the self-beating (homodyne) mode, can be written as: [43,44]:

$$g_2(t) - 1 = \left[ A_1 \exp\left(\frac{t}{\tau_1}\right) + A_2 \exp\left(\frac{t}{\tau_2}\right) \right]^2 \qquad (1)$$

where $A_1$ and $A_2$ are the amplitudes of the two relaxation processes, $t$ is the "lag" (or "delay") time of the photon correlations and $\tau_1$ and $\tau_2$ are the characteristic relaxation times. For a diffusive process, the relaxation time $\tau$ is related to the diffusion coefficient, $D$, as follows [43,44]:

$$\tau = \frac{1}{Dq^2} \qquad (2)$$

where $q$ is the difference in the wave number between incident and scattered light, $q = \left(\frac{4\pi n}{\lambda}\right) \sin\left(\frac{\theta}{2}\right)$, $n$ is the refractive index of the solvent, $\lambda$ is the wavelength of the incident light in vacuum ($\lambda$ = 633 nm for our set-up) and $\theta$ is the scattering angle. For monodisperse, non-interacting, spherical Brownian particles the hydrodynamic radius $R$ can be calculated with the Stokes-Einstein relation [43,44]:

$$R = \frac{k_B T}{6\pi \eta D} \qquad (3)$$

where $k_B$ is Boltzmann's constant, $T$ is the temperature and $\eta$ is the shear viscosity of the medium.

3. **Results**

Figure 1 shows the liquid-solid phase diagram of TBA-water solutions at ambient pressure, as determined by Kasraian *et al.* [13]. TBA and water are completely miscible under ambient conditions. The dotted region in this figure denotes the approximate concentration and temperature range where micelle-like fluctuations (dynamic molecular clustering) and thermodynamic anomalies are reported for TBA-water solutions [7,21-27]. At this concentration and temperature range, when a small amount of a hydrophobe is added to TBA-water solutions (or a commercial TBA sample is contaminated by hydrophobic impurities) mesoscopic droplets are observed [7,34,35].

Figure 2 shows the intensity auto-correlation function obtained for a TBA-water-CHX solution (0.1 mass % CHX and 26 mass % TBA) at $T$ = 25 °C and scattering angle $\theta$ = 45° ($q$ = $10^7$ m$^{-1}$). Figure 2 shows the presence of two exponential relaxation processes – a fast process with a relaxation time of 80 μs and a slow process with a relaxation time of 22 ms. In addition, Figure 2 also shows the presence of a non-exponential tail of long-time relaxations. The fast process corresponds to the molecular diffusion, with a diffusion coefficient of $1.2 \times 10^{-6}$ cm$^2$/sec, which, in accordance to Eq. (3), corresponds to a hydrodynamic radius (correlation length) of about 0.7 nm. The slower process, with a relaxation time of 22 ms, also exhibits diffusive behavior (as shown for a similar system in our previous work [34]), with an average diffusion coefficient of $4.1 \times 10^{-9}$ cm$^2$/sec. This relaxation corresponds to the Brownian diffusion of mesoscopic droplets having an average hydrodynamic radius of about 200 nm, but also displaying some polydispersity [45]. It has been observed that as the temperature is lowered, the size of the droplets do not change significantly, but their number increases (as observed by an enhancement in the light scattering intensity) [34]. As the temperature is raised, these droplets disappear, but are observed again when the system is cooled [34]. These droplets are highly long-lived, stable over a year. We call this phenomenon, manifested by the slow mode in dynamic light scattering, the mesoscale solubilization. The long-time tail of the correlation function may be due to the presence of occasionally occurring large aggregates of the order of tens of microns [34].

In order to understand the mesoscale solubilization further, macroscopic and mesoscopic ternary phase behavior of three different systems, TBA-water-PO, TBA-water-IBA, and TBA-water-CHX, were studied. In each of these three systems, the hydrophobe is completely miscible with TBA, but exhibits partial miscibility with water. PO exhibits the smallest miscibility gap with water [41], CHX is almost immiscible with water [46], while the solubility of IBA in water is intermediate [47].

The ternary phase diagram of TBA-water-PO system is shown in Figure 3a. The region inside the binodal curve is the two-phase region, while the region outside the binodal curve is the macroscopically homogeneous one-phase region. Various samples within the one-phase region were prepared and analyzed by static and dynamic light scattering. The dotted area of the ternary phase diagram corresponds to a region where the light scattering intensity is at least an order of magnitude higher than the intensity observed for the corresponding binary systems. Additionally, the dynamic auto-correlation functions obtained from this region show the presence of mesoscopic droplets, with a hydrodynamic radius of about 100 nm. We attribute this phenomenon to the mesoscale solubilization.

Figures 3b and 3c show the TBA-water-IBA and TBA-water-CHX phase diagrams, respectively. The dotted areas in these figures correspond to mesoscale solubilization, where mesoscopic droplets, with a hydrodynamic radius of about 100 nm are detected by dynamic light scattering.

To further distinguish between mesoscale solubilization and molecular diffusion, intensity auto-correlation functions for three samples from the TBA-water-IBA system at $T$ = 10 °C and scattering angle $\theta$ = 45° ($q$ = $10^7$ m$^{-1}$) are shown in Figure 4. The concentrations and properties of the three samples are given in Table 1. The correlation functions from samples A and B show the presence of a single exponential relaxation process, corresponding to molecular diffusion. These samples exhibit a diffusion coefficient of $7.0 \times 10^{-8}$ cm$^2$/sec and $1.6 \times 10^{-7}$ cm$^2$/sec, corresponding to a hydrodynamic radius of 5 nm and 2 nm, respectively. Sample A is closer to the critical point of the solution and hence exhibits larger concentration fluctuations (manifested by the larger correlation length and slower diffusion). The intensity auto-correlation function from sample C shows the presence of an additional slow relaxation process, associated with the mesoscale solubilization. The mesoscopic droplets have an average hydrodynamic radius of about 130 nm and an average diffusion coefficient of $6.7 \times 10^{-12}$ cm$^2$/sec.

## 4. Discussion

We would like to clarify the ambiguity associated with the definition of solubility and differentiate between molecular solubility, mesoscale solubilization, and macrophase separation. Although molecular solubility and macrophase separation are thermodynamically well-defined stable states, the role of mesoscale solubilization in the thermodynamic stability of the system is still an open question.

Molecular solubility of nonpolar solutes (hydrophobes) in water can be explained by the phenomenon of hydrophobic hydration [48,49]. Water molecules surrounding a nonpolar solute form a hydrogen-bonded shell around the solute molecule. This shell is similar to a clathrate shell and the water molecules in the shell do not interact strongly with the nonpolar solute in the core [50]. However, molecular solubility in aqueous solutions of nonionic hydrotropes is quite different, driven by strong interactions between solute and water molecules.

Molecular solubility in aqueous solutions of nonionic hydrotropes can be explained by the concept of clustering. Clustering refers to the formation of transient hydrogen bonds between water molecules and the polar groups of hydrotropes, resulting in micelle-like structural fluctuations. In our previous publication [7], we have confirmed the formation of such short-ranged (<1nm), short-lived (tens of picoseconds) clusters in aqueous solutions of TBA. Such clusters are formed due to strong hydrogen bonds between the water molecules and the polar groups of the solute molecules. The hydrogen-bonded network forms a shell structure, which surrounds the nonpolar parts of the solute molecules. These clusters create transient local domains of polar and nonpolar regions and lead to an effective solubilization of the water-hydrotrope mixtures. Although clustering in aqueous solutions of hydrotropes is a dynamic phenomenon, the molecular solubility affected by this clustering is a concept of thermodynamic equilibrium.

Another misleading concept in the literature is the distinction between the above-mentioned clusters and concentration fluctuations. The micelle-like clusters may have the same length scale as the correlation length of the concentration fluctuations when far away from the critical point, but very different dynamics. Clusters do not relax by diffusion, but by the reorientation of hydrogen bonds. They are too fast to be detected by dynamic light scattering (whose time domain is hundreds of nanoseconds and greater), and have too large a wavenumber to be detected by static light scattering. On the other hand, the correlation length of concentration fluctuations corresponds to the hydrodynamic radius obtained from molecular diffusion (as shown in the previous section for TBA-water-IBA systems). Molecular diffusion in aqueous hydrotrope solutions has a time scale of about tens of microseconds at $q \sim 10^7$ m$^{-1}$ (far away from the critical point), when probed by dynamic light scattering. The corresponding hydrodynamic radius (correlation length) is of the same order as the length scale of the clusters (< 1 nm). However, when close to the critical point the relaxation time scale could increase up to tens of milliseconds (at $q \sim 10^7$ m$^{-1}$) and the associated length scale, which is the correlation length of the concentration fluctuations, could reach hundreds of nanometers. Such critical concentration fluctuations, which show up close to a second order phase transition, can be detected by both static and dynamic light scattering [43,51].

Next, we discuss the phenomenon of the mesoscale solubilization. We explain the occurrence of mesoscopic droplets (with a length scale of about 100 nm) by the mesoscale solubilization of a hydrophobic compound in aqueous hydrotrope solutions at the temperature and concentration range where structural fluctuations are seen in the binary hydrotrope-water solution. Recent molecular dynamics simulations have shown that the hydrophobic compound stabilizes the structural fluctuations occurring in the binary hydrotrope-water solution, resulting in the formation of larger aggregates (2 to 3 nm), which are stable during the lifetime of the simulations (50 ns) [52]. These aggregates consist of a hydrophobe-rich core surrounded by a hydrogen bonded water-hydrotrope shell. We speculate that the mesoscopic droplets also have a hydrophobe-rich core surrounded by a hydrogen bonded water-hydrotrope shell. These droplets have been observed to be highly stable, without any significant change in their size or polydispersity, over long periods of time, from a few months (as in TBA-water-PO system) to over a year (as in TBA-water-CHX system) [45].

In order to understand the role of the mesoscopic droplets in the thermodynamic stability of the system, we investigated various samples in the two-phase region of the TBA-water-CHX ternary system, when the system has macroscopically phase separated. A ternary sample (whose overall concentration is 13 mass % TBA and 29 mass % CHX) was monitored over a period of three months. Figure 5 shows the auto-correlation functions (obtained from dynamic light scattering) observed in each of the two phases after 3 days of sample preparation. As seen from the figure, the organic layer shows no mesoscopic droplets, while the aqueous layer shows the presence of mesoscopic droplets of the same size and similar polydispersity as in the macroscopically homogeneous samples [45]. However when monitored over time, it was observed that the light scattering intensity decreased and after three months no mesoscopic droplets were detected in either of the phases. The liquid-liquid interface also did not show any unusual behavior. This indicates that in a two-phase macroscopically separated system, as long as there is an "infinite" reservoir of the hydrophobe-rich phase, the mesoscopic droplets will eventually breakdown, with their constituents joining each of the phases. However, the mesoscopic droplets observed in the macroscopically homogeneous one-phase region are much more stable, remaining unchanged over a period of a year.

The above-mentioned results lead to an important question: whether mesoscale solubilization is a kinetically arrested event or a thermodynamically equilibrium phenomenon? It appears that under certain circumstances, mesoscale solubilization is an equilibrium phenomenon, but with a very shallow minimum in the free energy, as seen in TBA-water-CHX system where the droplets remain almost unchanged over a period of a year [45]. The disappearance of these droplets in the presence of an "infinite" reservoir of CHX could be an indication that the droplets have condensed on to the interface. In other cases, the mesoscopic droplets could be nonequilibrium, but with very slow kinetics (as observed in TBA-water-PO systems [45]), possibly due to a high-energy barrier. In such a case, the mesoscopic droplets, if given enough time, could eventually aggregate to form a macroscopic phase. These two cases are difficult to discriminate. Practically speaking, thermodynamic solubility could be indistinguishable from a kinetically arrested state if the kinetics is extremely slow.

The reasons for the long-term stability of the mesoscopic droplets in the macroscopically homogeneous one-phase region of the TBA-water-CHX system may be two-fold. The strong hydrogen bonds between water and TBA molecules, and the tendency of TBA to orient itself "perfectly" between water and oil could shield the hydrophobic core of the mesoscopic droplets from the aqueous solution, thus lowering the effective oil-water surface tension [53]. The shield may also act as a relatively "rigid" membrane, resisting deformation to produce smaller droplets [54,55]. Another reason for the high stability of the droplets could be such that in the absence of a large reservoir of the CHX-rich phase, the hydrophobes present in the core of the droplets do not have enough driving force to overcome the hydrogen-bonded shield and reach the interface [56]. Hence, under these conditions, the mesoscopic droplets could remain almost unchanged for very long periods of time, leading to a practically stable system.

However, there is another interesting phenomenon that was observed at the liquid-liquid interface of many hydrotrope-water-hydrophobe systems [57]. As an example, the interface of one of the TBA-water-CHX samples (whose overall concentrations are 16 mass % TBA and 40 mass % CHX) is shown in Figure 6. After preparing the sample and equilibrating it for about 24 hours, a new soap-like "phase" was observed at the oil-water interface. This new "phase" seemed fluid-like, without any specific adhesion to the walls of the vial. Collecting this "phase" from the macroscopic liquid-liquid interface by a micropipette turned out to be unsuccessful, as this phase was easily disturbed and easily disintegrated. We speculate that this novel "phase" may form due to the presence of additional impurities or due to the surface activity of TBA [53], which could attract the mesoscopic droplets towards the water-oil interface. In order to verify this hypothesis, we prepared a ternary TBA-water-CHX sample, by using a different source of TBA, one with lower purity (0.997+) than was used for previous sample preparations. When this sample was monitored over a few hours after preparation, the aqueous-rich phase showed the presence of mesoscopic droplets, while the organic-rich phase did not show any droplets. In addition, the interface of this sample showed the presence of the novel "phase".

5. **Conclusions**

In this work, we have addressed the ambiguity behind the definition of solubility, differentiating between molecular solubility, mesoscale solubilization, and macro-phase separation. Molecular solubility in aqueous solutions of nonionic hydrotropes can be associated with molecular clustering. Clustering exhibits non-diffusive behavior, making and breaking of transient non-covalent bonds between water and hydrotrope molecules. Clustering has a length scale of about a nanometer and a lifetime of about tens of picoseconds. The length scale of the clusters may be similar to the length scale of the concentration fluctuations (when far away from the critical point), which corresponds to the hydrodynamic radius (correlation length) obtained from molecular diffusion. However the time-scale of the clusters is much faster than the time scale of molecular diffusion (tens of picoseconds for the former vs. tens of microseconds for the latter).

In the presence of a hydrophobic compound, the structural fluctuations occurring in binary water-hydrotrope solutions seem to be stabilized and result in the formation of mesoscopic droplets. We call this phenomenon the mesoscale solubilization. Mesoscale solubilization is not unique to the systems studied in this work. Mesoscale solubilization is observed in a variety of other aqueous systems, such as in 3-methylpyridine, tetrahydrofuran, 2-butoxyethanol, isobutyric acid [34-40,57-61]. Mesoscale solubilization leads to the creation of a novel kind of practically stable colloids, made from small molecules without the addition of surfactants, polymers, emulsifiers, or charged species. Mesoscale solubilization has a variety of applications from encapsulated drug delivery to design and development of various pharmaceutical and cosmetic products [62-63].

Unique features in the phenomenon of mesoscale solubilization are the size of the mesoscopic droplets and their stability. The droplets have a characteristic size, of the order of 100 nm. This size does not seem to depend on the type of the hydrophobe or the hydrotrope, nor does it seem to strongly depend on temperature. However, the number of these particles strongly increases as the temperature is lowered. A quantitative formulation of the conditions for thermodynamically stable mesoscale solubilization versus a kinetically arrested state requires further experimental and theoretical studies.

**Funding Sources**

This research is supported by the Division of Chemistry (Grant No. CHE-1012052) and the Division of Industrial Innovation and Partnerships (Grant No. IIP-1261886) of the National Science Foundation.

**Acknowledgement**

We appreciate useful discussions with P. J. Collings, D. Frenkel, J. B. Klauda, J. Leys, V. Molinero, A. Onuki, M. Sedlák, M. Schick, and W. Schröer. We also thank E. Altabet, Y. Alsaid, S. Hayward, S. Kim, and D. Weglein for helping us with sample preparation and light scattering experiments.

**Table 1:**

Concentrations of TBA-water-IBA samples studied in this work. Their correlation functions are shown in Figure 4.

| Sample # | Mass Fr. TBA | Mass Fr. Water | Mass Fr. IBA | Refractive Index | Kinematic Viscosity at 10 °C ($\times 10^{-6}$ m$^2$/s) |
|---|---|---|---|---|---|
| A | 0.29 | 0.60 | 0.11 | 1.3691 | 7.60 |
| B | 0.26 | 0.65 | 0.09 | 1.3650 | 6.36 |
| C | 0.10 | 0.86 | 0.04 | 1.3559 | 2.96 |

**List of Figures and Figure Captions:**

**Figure 1:** Solid-liquid phase diagram (filled black circles) of TBA-water solutions at ambient conditions (based on experimental data from ref. [13]). The dotted area schematically shows the region where molecular clustering and thermodynamic anomalies are reported. This area also corresponds to the temperature and concentration region where mesoscale solubilization is observed in TBA-water-hydrophobe solutions. The curve joining the filled black circles is a guide to the eye. The horizontal and vertical lines separate the different phases of the TBA-water hydrate.

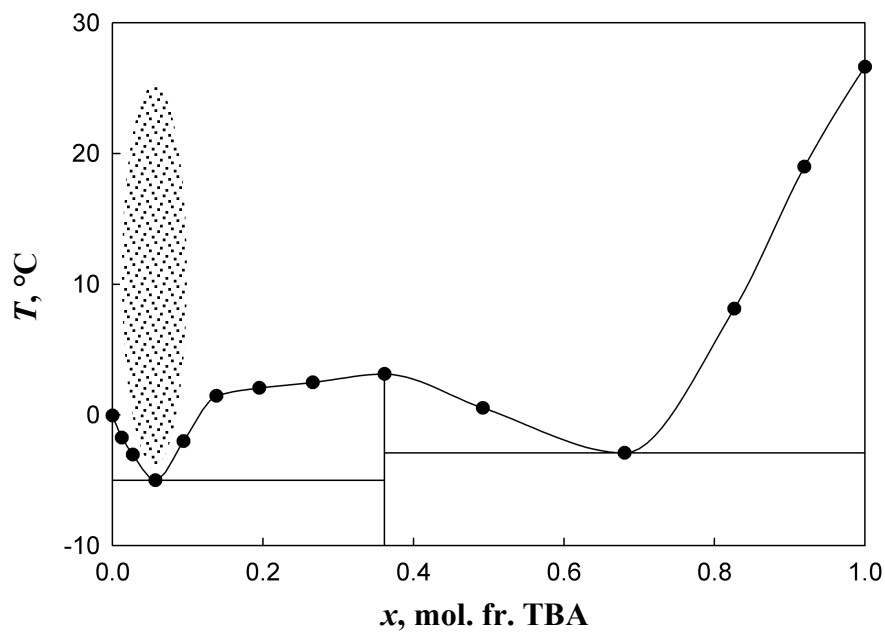

**Figure 2:** Intensity auto-correlation function for a TBA-water-CHX solution (0.1 mass % CHX and 26 mass % TBA) at $T$ = 25 °C and a scattering angle of $\theta$ = 45°. The black line is a fit to Eq. (1). The correlation function shows the presence of two exponential relaxation processes. The fast process, with a relaxation time of 80 μs and a slow process with a relaxation time of 22 ms. The fast process corresponds to molecular diffusion with a diffusion coefficient of $1.2 \times 10^{-6}$ cm$^2$/sec and a hydrodynamic radius of 0.7 nm. The slow process, which we call mesoscale solubilization, corresponds to long-lived, highly stable mesoscopic droplets of about 200 nm in size, with an average diffusion coefficient of $4.1 \times 10^{-9}$ cm$^2$/sec.

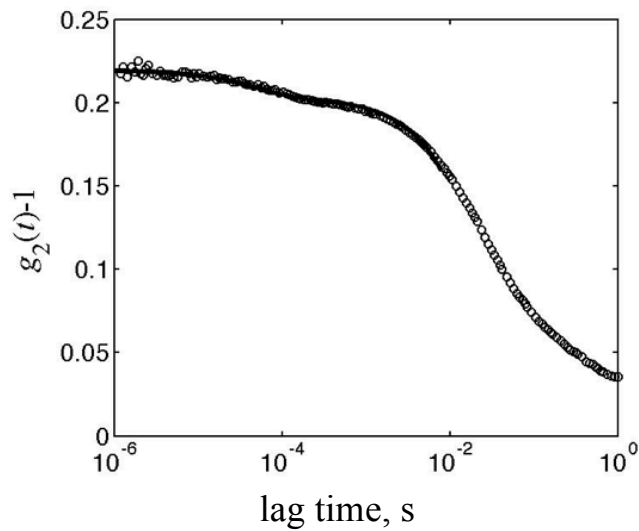

**Figure 3:** Figures 3a, 3b, and 3c represent ternary phase diagrams of TBA-water-hydrophobe systems at ambient conditions. All concentrations are in mass fractions. The smooth line across the points is a guide to the eye. Open circles represents approximate location of the critical point. The dashed line from the vertex of the hydrophobe represents concentrations with a constant TBA-Water ratio (25:75 mass basis / 7:93 mole basis) where thermodynamic anomalies in the binary TBA-water solution exhibit extrema. The dotted area in the phase diagram, characterized as mesoscale solubilization, shows the region where mesoscopic droplets are observed.

a) TBA-Water-PO phase diagram ($T \cong 25$ °C)

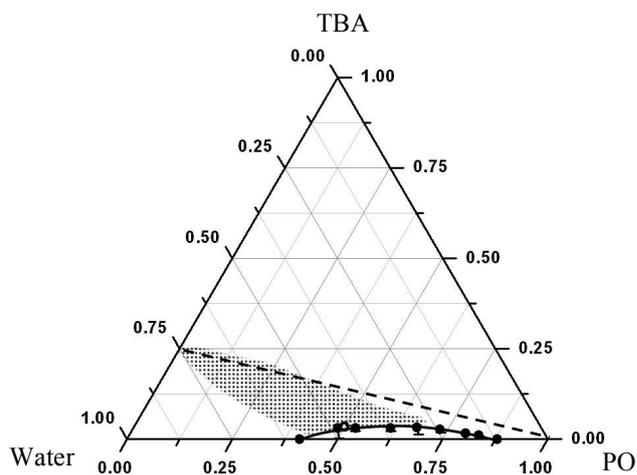

b) TBA-Water-IBA phase diagram ($T \cong 21$ °C)

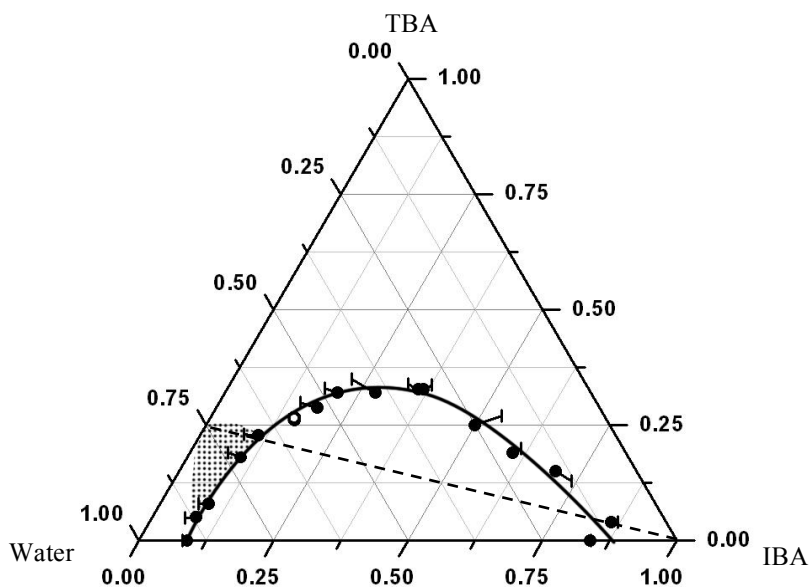

c) TBA-Water-CHX phase diagram ($T \cong 21\ °C$)

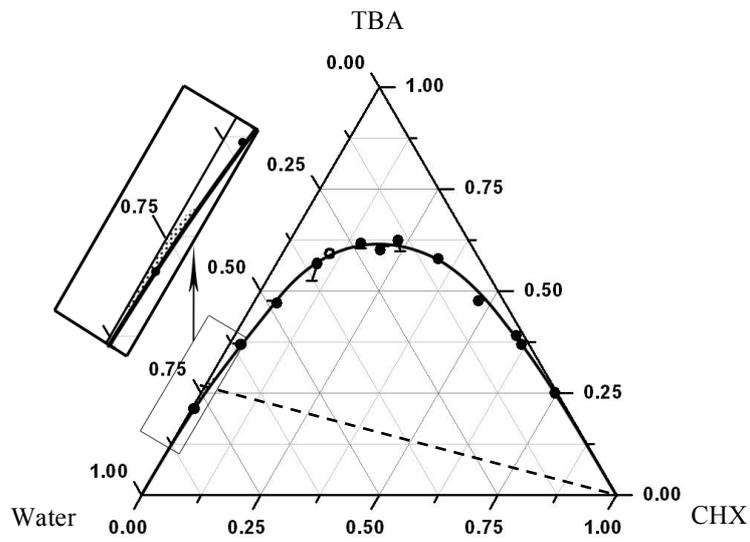

**Figure 4:** Intensity auto-correlation functions for TBA-water-IBA solutions at $T$ = 10 °C and a scattering angle of $\theta$ = 45°. Sample concentrations are displayed in Table 1. Squares (sample # A) and crosses (sample # B) represent solutions that are close to the critical point of the system, indicating a relaxation time of about 1.3 ms and 0.6 ms respectively. These correspond to correlation lengths of 5 nm and 2 nm respectively. The circles correspond to sample C from Table 1, which additionally show the presence of mesoscopic droplets. The droplets have a relaxation time of about 37 ms, which corresponds to a length scale of about 130 nm.

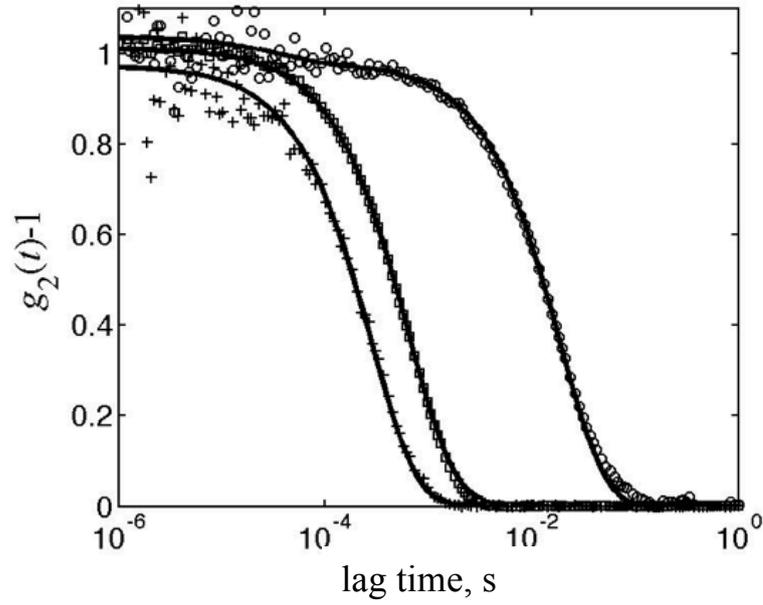

**Figure 5:** Intensity auto-correlation functions for a TBA-water-CHX system in the two-phase region measured after ~ 3 days of preparation ($T$ = 25 °C and scattering angle $\theta$ = 45°). Overall sample concentration is 13 mass % CHX and 29 mass % TBA. Correlation function from the aqueous phase (circles) shows the presence of mesoscopic droplets, while the correlation function from the CHX rich phase (crosses) shows no such phenomenon. When measured after a period of 3 months, the aqueous phase and the CHX rich phase show no correlations, thus indicating that in a two-phase system, the mesoscopic droplets phase separate, albeit very slowly.

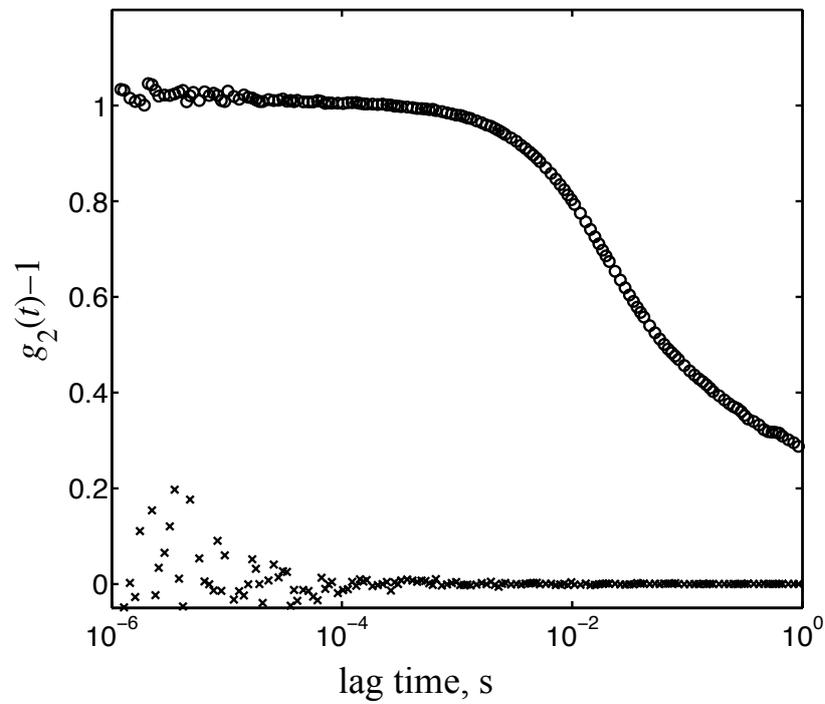

**Figure 6:** Image of TBA-water-CHX system in the two-phase region. The vial diameter is 2.5 cm. The overall concentration of the sample is 16 mass % CHX and 40 mass % TBA. The sample shows the presence of a novel phase at the interface of the aqueous rich and CHX rich layers.

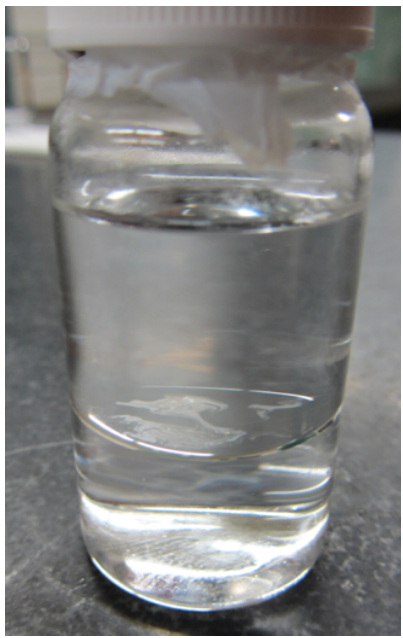